\documentclass[%
 aip,
 amsmath,amssymb,
 reprint,%
]{revtex4-1}

\usepackage{graphicx}% Include figure files
\usepackage{dcolumn}% Align table columns on decimal point
\usepackage{bm}% bold math
\usepackage[utf8]{inputenc}
\usepackage[T1]{fontenc}
\usepackage{mathptmx}

\newcommand{\la}{\lambda}
\newcommand{\ga}{\gamma}

\newcommand{\al}{\alpha}
\newcommand{\om}{\omega}

\newcommand{\oq}{\overline{q}}

\newcommand{\tk}{\widetilde{k}}
\newcommand{\tom}{\widetilde{\omega}}

\newcommand{\prt}{\partial}

\begin{document}

%\preprint{AIP/123-QED}

\title{Asymptotic theory of not completely integrable soliton equations}

\author{ A. M. Kamchatnov}

\affiliation{Institute of Spectroscopy,
Russian Academy of Sciences, Troitsk, Moscow, 108840, Russia}
%\altaffiliation[Also at ]{Moscow Institute of Physics and Technology, Institutsky
%lane 9, Dolgoprudny, Moscow region, 141700, Russia.}%Lines break automatically or can be forced with \\

\email{kamch@isan.troitsk.ru.}

\date{\today}

\begin{abstract}
We develop the theory of transformation of intensive initial nonlinear wave pulses to
trains of solitons emerging at asymptotically large time of evolution. Our approach is
based on the theory of dispersive shock waves in which the number of nonlinear oscillations
in the shock becomes the number of solitons at the asymptotic state. We show that this number
of oscillations, which is proportional to the classical action of particles associated with the
small-amplitude edges of shocks, is preserved by the dispersionless flow. Then the
Poincar\'e-Cartan integral invariant is also constant and therefore it reduces to the quantization
rule similar to the Bohr-Sommerfeld quantization rule for linear spectral problem associated with
completely integrable equations. This rule yields a set of `eigenvalues' which are related with
the asymptotic solitons' velocities and other their characteristics. Our analytical results agree
very well with the results of numerical solutions of the generalized
nonlinear Schr\"odinger equation.
\end{abstract}


% 47.35.Jk 	Wave breaking
% 02.30.Ik 	Integrable systems
% 02.30.Jr 	Partial differential equations
% 47.35.Fg 	Solitary waves

\maketitle

\begin{quotation}
In many typical situations an intensive enough initial wave pulse evolves eventually to some number 
of solitons, and therefore it is very important to be able to predict velocities of these asymptotic 
solitons and other their characteristics. If the nonlinear wave equation under consideration is completely 
integrable, then this problem can be solved by finding the eigenvalues of the linear spectral
problem associated with this equation. The theory becomes especially effective when the number
of solitons is large, so that the quasiclassical asymptotical method can be applied to the 
spectral problem. However, in case of not completely integrable equations a similar approach
was only known for a special class of initial conditions when they correspond to simple waves
of the dispersionless approximation. In this paper, we generalize this theory to arbitrary 
smooth enough initial pulses described by two wave variables. We obtain the generalized
Bohr-Sommerfeld quantization rule which defines a set of `eigenvalues' corresponding to given
initial conditions and show how these eigenvalues are related with physical parameters of
asymptotic solitons. The theory agrees very well with numerical simulations and sheds new 
light on the quasiclassical limit of completely integrable equations.
\end{quotation}

\section{Introduction}\label{intro}

If a nonlinear wave system supports propagation of solitons, then an intensive
enough initial pulse evolves eventually into a certain number $N$ of solitons and some
amount of linear radiation. In view of universality of this phenomenon, it is very important
to be able to predict parameters of emerging solitons (e.g., their velocities and amplitudes)
from given initial distributions of the wave variables. This problem admits a
straightforward solution in case of completely integrable nonlinear wave equations;
see, e.g., Refs.~\cite{nmpz-80,as-81,newell-85} and references therein. Such equations are
associated with some linear spectral problems where the wave distributions play the
role of `potentials'. Though these `potentials' evolve with time according to the wave
equation under consideration, the spectrum of the associated linear problem does not change,
i.e. the evolution is isospectral, and each discrete eigenvalue corresponds to a certain
soliton at the asymptotic stage of evolution at $t\to\infty$. Hence, if we find the
spectrum of the linear problem for given initial wave distributions, then the discrete
eigenvalues provide all necessary information about solitons' parameters at asymptotically
large time $t$.

This theory considerably simplifies if the final number of solitons is large, $N\gg1$,
and the initial distributions are smooth enough, so we can apply the quasiclassical WKB
method to the linear spectral problem. For example, in case of Korteweg-de Vries (KdV)
equation associated with the stationary Schr\"{o}dinger equation \cite{ggkm-67}, the
well-known Bohr-Sommerfeld quantization rule readily gives approximate values of the
discrete eigenvalues very simply related with solitons' velocities and amplitudes
\cite{karpman-67,karpman-73}. As usual in the WKB method, the asymptotic formulas give quite
accurate results even for small quantum numbers. Similar Bohr-Sommerfeld quantization
rule was derived in Refs.~\cite{JLML-99,kku-02} for the Zakharov-Shabat spectral
problem associated with nonlinear Schr\"{o}dinger (NLS) equation \cite{zs-73}. However,
when the nonlinear wave equation under consideration is not completely integrable, such a
direct approach becomes impossible.

To find a more general method for calculation of parameters of asymptotic solitons, we need
to consider in some more detail the process of transformation of an initially smooth pulse
to a sequence of solitons. One can distinguish in this process three typical stages.
At first, the pulse's profile gradually steepens due to nonlinear effects, but it keeps a
smooth enough form without any oscillations, so its evolution obeys with high accuracy the
dispersionless (hydrodynamic) limit of nonlinear wave equations under consideration.
When the profile's slope becomes very large, the dispersion effects start to influence
on the evolution of the wave, so oscillations are generated after the so-called wave breaking
moment of time. Such regions of nonlinear oscillations are called dispersive shock waves (DSWs)
and, according to Gurevich and Pitaevskii \cite{gp-73}, they can be approximated by modulated
periodic solutions of the nonlinear wave equations. Evolution of the modulation parameters
obeys the Whitham modulation equations \cite{whitham,Whitham-74}. In Gurevich-Pitaevskii
approximation, a DSW occupies a finite expanding region of space. At one edge of the DSW
solitons are gradually formed and the opposite small-amplitude edge propagates with some group
velocity along a smooth part of the evolving pulse. Thus, at the second stage of evolution the
wave structure consists of one or two DSWs adjoined with smooth parts of the pulse and the
DSWs expand gradually over the whole pulse. Numbers of oscillations in DSWs increase with time
due to difference between phase and group velocities of linear waves at the small-amplitude
edges where oscillations enter into the DSWs regions \cite{gp-87}. At the last third stage
of evolution of an initially localized pulse with a finite length its smooth parts disappear
and the number of oscillations in DSWs is stabilized, so the solitons acquire their asymptotic
values of parameters as $t\to\infty$, when distances between solitons become much greater than
their widths. Thus, asymptotic solitons form gradually from small-amplitude oscillations
which enter into the DSW region at its small-amplitude edge, so we have to consider propagation
of this edge in some more detail.

Let the nonlinear wave dynamics under consideration be described by two variables which for
definiteness we shall call ``density'' $\rho$ and ``flow velocity'' $u$. The periodic solutions
$\rho=\rho(x,t), u=u(x,t)$ are periodic functions of the phase $\theta$ which in non-modulated case
has the form $\theta=kx-\om t=k(x-Vt)$, where $V$ is the phase velocity of the traveling wave.
In a slightly modulated wave, the wave number $k$ and the frequency $\om$ become slow functions of
$x$ and $t$. Since locally the wave can still be considered in the leading approximation as uniform,
they can be defined as
\begin{equation}\label{eq1}
  k=\theta_x,\qquad \om=-\theta_t
\end{equation}
for some phase $\theta(x,t)$, and then one of the modulation equations can be written in the
form of `the number of waves' conservation law \cite{whitham,Whitham-74}
\begin{equation}\label{eq2}
  k_t+\om_x=0,
\end{equation}
where $k$ plays the role of density of waves and $\om$ is their flux.

At the small-amplitude edge we can distinguish the short wavelength oscillation at the length scale
$\sim2\pi/k$ and the slowly changing ``background distributions'' $\rho_b(x,t),u_b(x,t)$ which
change at the length scale of the initial pulse length $\sim l$. Hence we can write
$\rho(x,t)=\rho_b(x,t)+\rho'(x,t),u(x,t)=u_b(x,t)+u'(x,t)$ where the smooth functions
$\rho_b(x,t),u_b(x,t)$ obey the motion equation in the dispersionless (hydrodynamic) limit
obtained by neglecting term with higher order space and/or time derivatives, and the small-amplitude
oscillations $\rho'(x,t),u'(x,t)$ obey the equations linearized with respect to deviations from
the background flow. Since at the length scale $\sim2\pi/k$ the background flow can be considered as
constant, we obtain in the main approximation from the linearized equations with
constant coefficients the harmonic wave solutions $\rho',u'\propto\exp[i(kx-\om t)]$ with the
dispersion relation $\om=\om(k,\rho_b,u_b)$. A slightly modulated linear wave packet propagating
along a slowly changing background still consists of harmonics with the same dispersion law
which satisfies the number of waves conservation law (\ref{eq2}), but $\rho_b,u_b$ become in this
case the slowly changing solutions $\rho=\rho(x,t),u=u(x,t)$ of the dispersionless equations
(see Ref.~\cite{Whitham-74}). In Gurevich-Pitaevskii theory~\cite{gp-73} of DSWs, the small-amplitude
edge is represented by such a linear wave packet and according to the conservation law (\ref{eq2})
oscillations enter into the DSW region where their amplitudes grow from zero at the small-amplitude
edge to soliton-like propagation at the opposite soliton edge. Since the small-amplitude edge
propagates with the group velocity $v_g=\prt\om(k,\rho,u)/\prt k$, along its path the flux is
Doppler-shifted, so the number of oscillations inside the DSW region changes with time as
(see Refs.~\cite{gp-87,kamch-21a})
\begin{equation}\label{eq3}
  \frac{dN}{dt}=\frac1{2\pi}\left|k\frac{\prt\om}{\prt k}-\om\right|.
\end{equation}
The expression in the right-hand side coincides, up to a constant factor, with Lagrangian of
a point particle associated with the small-amplitude edge of the DSW, if in accordance with the
well-known optics-mechanical analogy (see, e.g., Ref.~\cite{lanczos}) we identify $k$ and $\om$
with the particle's momentum and Hamiltonian, correspondingly. Then the number $N$ of solitons
formed from an intensive initial pulse is equal to the mechanical action produced by an
associated with the packet point-like particle,
\begin{equation}\label{eq4}
  N=\frac{S}{2\pi}=\frac1{2\pi}\int(kdx-\om dt),
\end{equation}
where $dx=v_gdt$ and integration is taken over the packet's path.
If the initial distributions $\rho_0(x),u_0(x)$ correspond to unidirectional simple wave propagation
of the background pulse, that is there exists a functional dependence between $\rho$ and $u$
corresponding to a constant value of one of the Riemann invariants of the dispersionless
equations~\cite{LL-6}, then the wave number $k$ can be expressed as a function of the value of
$\rho$ at the point where the small-amplitude edge is located at this moment of time, $k=k(\rho)$.
Then we have also $\om=\om(k,\rho)$, where $\rho=\rho(x,t)$ is a solution of the Hopf equation
\begin{equation}\label{eq5}
  \rho_t+V_0(\rho)\rho_x=0
\end{equation}
to which the dispersionless equations can be reduced, $k=k(\rho)$ is the solution of the equation
\begin{equation}\label{eq6}
  \frac{dk}{d\rho}=\frac{\prt\om/\prt\rho}{V_0-\prt\om/\prt k}
\end{equation}
with the initial condition
\begin{equation}\label{eq7}
  k(0)=0,
\end{equation}
where it is assumed that wave breaking occurs at the point with $\rho=0$. Equation (\ref{eq6})
was first obtained in Ref.~\cite{el-05} by means of analysis of Whitham equations for DSWs at their
small-amplitude edge and then it was derived from the Hamilton equations for propagation of
linear wave packets along simple wave pulses in Refs.~\cite{kamch-20a,ks-21}. When the function
$k=k(\rho)$ is found, then, as was shown in Ref.~\cite{kamch-19}, one can find the path $x=x(t)$
of the small-amplitude edge of a DSW and, consequently, calculate the integral in Eq.~(\ref{eq4})
for some particular nonlinear wave equations (see Refs.~\cite{kamch-20a,kamch-21,cbk-21}).
In this case the final result can be written in the form
\begin{equation}\label{eq8}
  N=\frac1{\pi}\int_{-\infty}^{\infty} k(\rho_0(x))dx,
\end{equation}
where $\rho_0(x)$ is the initial distribution of $\rho$. Such a formula was suggested earlier
in Refs.~\cite{egkkk-07,egs-08} under supposition that one can extend solution of the Whitham
equations to the unmodulated yet part of the pulse in such a way, that the total number of
oscillations in DSW and unmodulated part remains constant during evolution of the pulse.
Generalization of this rule to more general solutions $k=k(\rho,\oq)$, $\oq$ is an integration constant
in a solution of Eq.~(\ref{eq6}), yields also parameters of solitons at the asymptotic stage of 
evolution \cite{egs-08}. This theory agrees with numerical solutions of nonlinear wave equations 
and with experimental results obtained in Ref.~\cite{mfweh-20} for a viscous fluid conduit.

So far this theory was limited to an important but particular case of an initial simple-wave
pulse evolving eventually to a number of solitons and the aim of this paper is to generalize
this approach to the general case of nonlinear wave evolutions describes by two variables
$\rho(x,t),u(x,t)$ and admitting solitonic propagation of waves. To this end, we notice that
equality of two expressions (\ref{eq4}) and (\ref{eq8}) can be interpreted as a special case
of preservation of a constant value of the Poincar\'{e}-Cartan integral invariants
\cite{poincare,cartan} for a mechanical system with the phase space $(x,k)$ and Hamiltonian
$\om(k,x,t)$. However, in this special case the tube of trajectories preserving the invariant
is defined by solutions of Eq.~(\ref{eq5}) rather than by solutions of the Hamilton equations
under consideration. In this paper, we extend this definition of Poincar\'{e}-Cartan integral invariants
to general hydrodynamic flows for two variables $\rho,u$ not restricted by the condition
that one of the Riemann invariants is constant and arrive at generalization of formula (\ref{eq8})
\begin{equation}\label{eq9}
  N=\frac1{\pi}\int_{-\infty}^{\infty} k(\rho_0(x),u_0(x),q_N)dx,
\end{equation}
where $\rho_0(x),u_0(x)$ are the initial distributions of these two variables and $q_N$ is an
integration constant in the solution of equations which define the function $k=k(\rho,u)$ of
two variables. One can say that $q_N$ corresponds to the $N$-th soliton in the asymptotic
distribution of solitons. Then if we decrease $N$ by one, then we get the value $q_{N-1}$
corresponding to the $(N-1)$-th soliton, and in this way we relate each $n$-th soliton
with the corresponding value $q_n$ of the generalized `Bohr-Sommerfeld quantization' rule (\ref{eq9}),
which extends the above mentioned method \cite{karpman-67,karpman-73,JLML-99,kku-02} of finding
solitons parameters for completely integrable wave equations to not completely integrable equations.
Relationship of the parameters $q_n$ with such physical parameters of solitons as, for example,
their velocities is established by means of the Stokes remark \cite{lamb, stokes} (see also
\cite{ai-77,dkn-03,kamch-20a}) that linear waves and exponentially small soliton tails obey the
same linearized equations, so the expressions for their phase or soliton velocities are
transformed to each other by the replacement $k\leftrightarrow-i\tk$, where $\tk$ is the
inverse half-width of a soliton.

We illustrate the formulated here approach by its application to the generalized NLS (gNLS)
equation and confirm its accuracy by comparison with numerical solutions.

\section{Poincar\'e-Cartan integral invariant and Bohr-Sommerfeld quantization rule}

Let the hydrodynamic (dispersionless) equations for the background flow evolution
can be written in the form of equations of the compressible fluid dynamics,
\begin{equation}\label{eq10}
\rho_t + (\rho u)_x = 0, \qquad u_t + uu_x+\frac{c^2}{\rho}\rho_x = 0,
\end{equation}
where $c=c(\rho)$ is the `sound velocity' related with the density $\rho$ by means of equation
of state $p=p(\rho)$, $c^2=dp/d\rho$, $p$ is the pressure. The characteristic velocities
of this system,
\begin{equation}\label{eq11}
  v_+=u+c,\qquad v_-=u-c,
\end{equation}
correspond to the sound waves propagating upstream and downstream the background flow with velocity $u$.
Equations (\ref{eq10}) can be transformed to the Riemann diagonal form
\begin{equation}\label{eq12}
%\begin{split}
\frac{\prt r_+ }{\prt t} + v_+ \frac{\prt r_+}{\prt x}=0,\quad
\frac{\prt r_- }{\prt t} + v_- \frac{\prt r_-}{\prt x}=0,
%\end{split}
\end{equation}
for the variables
\begin{equation}\label{eq13}
  r_{\pm}=\frac{u}{2}\pm\frac12\int_0^{\rho}\frac{cd\rho}{\rho}
\end{equation}
called Riemann invariants. If we substitute here the function $\rho=\rho(c)$,
then we get the Riemann invariants in the form
\begin{equation}\label{eq14}
  r_{\pm}=\frac{u}{2}\pm\sigma(c),\qquad \sigma(c)=\frac12\int_0^c\frac{c}{\rho(c)}\frac{d\rho}{dc}dc,
\end{equation}
so that $c$ can be used as a wave variable instead of $\rho$.
If Eqs.~(\ref{eq12}) are solved, then the physical variables $u,\rho,c$ can be expressed in terms
of the Riemann invariants,
\begin{equation}\label{eq15}
  u=r_++r_-,\quad \rho=\rho(r_+,r_-),\quad c=c(r_+,r_-).
\end{equation}

In case of unidirectional propagation of a background pulse we get a simple wave solution with one 
of the Riemann invariants constant. Let such a pulse propagate to the right through the uniform
quiescent medium with constant values of $\rho=\rho_R$, $c=c_R$, $u=u_R=0$. Then along this
pulse the variables $u$ and $c$ are related by the formula $u/2-\sigma(c)=-\sigma(c_R)$, so
that $c=c(u)$ and $r_+=u+\sigma(c_R)$, $v_+=u+c(u)\equiv V_0(u)$. Consequently, the second
equation (\ref{eq12}) is fulfilled identically and the first one reduces to the Hopf equation
\begin{equation}\label{eq16}
  \frac{\prt u}{\prt t}+V_0(u) \frac{\prt u}{\prt x}=0.
\end{equation}

We are interested in propagation of the small-amplitude edge of a DSW and we identify this
propagation with motion of a point particle with coordinate $x(t)$ and momentum $k(t)$ which
satisfy the Hamilton equations
\begin{equation}\label{eq17}
\frac{d x}{d t} = \frac{\prt \omega }{\prt k},\qquad
\frac{d k}{d t} = -\frac{\prt \omega }{\prt x}.
\end{equation}
In usual situations the Hamiltonian (dispersion relation) $\om$ is a function of $k,x,t$,
$\om=\om(k,x,t)$. Following the excellent textbook \cite{gant-66}, let us define in
the three-dimensional extended phase space $(x,k,t)$ an arbitrary closed curve
\begin{equation}\label{eq18}
  C_0=\left\{x=x_0(\eta),\,\,k=k_0(\eta),\,\,t=t_0(\eta)\,\,|\,\,0\leq\eta\leq1\right\},
\end{equation}
where the points corresponding to $\eta=0$ and $\eta=1$ coincide with each other. Then we can
calculate along this curve the integral
\begin{equation}\label{eq19}
  I_0=\oint_{C_0}(k\delta x-\om\delta t),
\end{equation}
where $\delta x$ and $\delta t$ denote differentials of the functions $x=x_0(\eta)$, $t=t_0(\eta)$
at the points of the contour $C_0$. Now, we define the flow in the phase space by the equations
\begin{equation}\label{eq20}
  \frac{dx}{dt}=Q(k,x,t),\qquad \frac{dk}{dt}=P(k,x,t)
\end{equation}
and, when solutions of these equations cross the contour $C_0$, we obtain a tube of trajectories
\begin{equation}\label{eq21}
  x=x(t,\eta),\qquad k=k(t,\eta),
\end{equation}
where $x(t_0(\eta),\eta)=x_0(\eta)$, $k(t_0(\eta),\eta)=k_0(\eta)$. If we take now another function
$t=t_1(\eta)$, then we get a new contour
\begin{equation}\label{eq22}
\begin{split}
  &C_1=\{x=x_1(\eta)=x(t_1(\eta),\eta),\\
  &k=k_1(\eta)=k(t_1(\eta),\eta),\,\,
  t=t_1(\eta)\,\,|\,\,0\leq\eta\leq1\}
  \end{split}
\end{equation}
and a new value of the integral similar to Eq.~(\ref{eq19}),
\begin{equation}\label{eq23}
  I_1=\oint_{C_1}(k\delta x-\om\delta t),
\end{equation}
which is taken over the contour $C_1$ around the tube of trajectories generated by the flow
(\ref{eq21}). Now one can ask, for which Eqs.~(\ref{eq20}) and corresponding flows (\ref{eq21})
the integrals (\ref{eq19}) and (\ref{eq23}) are equal to each other independently of the
choice of the contour $C_1$ (that is the choice of the function $t=t_1(\mu)$). As was shown
in Refs.~\cite{poincare,cartan,gant-66}, this is true for
\begin{equation}\label{eq24}
  Q=\frac{\prt \om}{\prt k},\qquad P=-\frac{\prt \om}{\prt x},
\end{equation}
that is when the flow obeys the Hamilton equations (\ref{eq17}). Thus, the invariance of the
Poincar\'e-Cartan integral
\begin{equation}\label{eq25}
  I=\oint_{C}(k\delta x-\om\delta t)
\end{equation}
means that the flow (\ref{eq20}) is Hamiltonian.

Now we change the perspective and  look at invariance of the integral (\ref{eq25}) from
a different point of view. As was indicated in Introduction, we are interested in invariance of
this integral with respect to flows generated by the hydrodynamic equations (\ref{eq10}) or
(\ref{eq16}). In this case the Hamiltonian $\om$ depends on $x$ and $t$ only via solutions
$\rho=\rho(x,t)$, $u=u(x,t)$ of these equations, i.e.,
\begin{equation}\label{eq26}
  \om=\om(k,\rho,u)
\end{equation}
and the flow is generated by the equation
\begin{equation}\label{eq27}
  \frac{dx}{dt}=u(x,t).
\end{equation}
Then invariance of the integral (\ref{eq25}) imposes certain conditions on the dependence
of the momentum (wave number) $k$ on the background variables,
\begin{equation}\label{eq28}
  k=k(\rho,u).
\end{equation}
If the condition of invariance is fulfilled, then the action (\ref{eq4}) can be reduced to the
generalized Bohr-Sommerfeld quantization rule (\ref{eq9}) which provides important information
about parameters of asymptotic solitons. We shall consider the condition of invariance of the
integral ({\ref{eq25}) separately for the simple wave flow (\ref{eq16}) and for the general
solutions of Eqs.~(\ref{eq10}),

\subsection{Simple wave background pulse}

In simple wave case, the small-amplitude edge propagates along the background pulse described by
a single variable $u=u(x,t)$ which evolution obeys the Hopf equation (\ref{eq16}). Then we have
$\om=\om(k,u)$ and we look for such a function $k=k(u)$, that the Poincar\'e-Cartan integral
(\ref{eq25}) remains the same for any contour $C$ around the tube generated by the flow 
$dx/dt=V_0(u)$. Following Ref.~\cite{gant-66}, we introduce a coordinate $\mu$ along the paths that
form the tube and assume that this coordinate is related with the flow by the relations
\begin{equation}\label{eq29}
  \frac{dx}{V_0(u)}=dt=\chi d\mu,
\end{equation}
where $\chi=\chi(t,x,k)$ is an arbitrary function: its choice fixes the choice of the coordinate
$\mu$ along the flow trajectories. Thus, for any fixed value of $\mu=\mathrm{const}$ we get a
point on every trajectory, so these points define the closed contour around the tube. As a result,
the integral (\ref{eq25}), generally speaking, becomes a function of $\mu$, but we want to find
such a function $k=k(u)$, that this integral does not depend on $\mu$ for any choice of $\chi$.
Differentiation of Eq.~(\ref{eq25}) with respect to $\mu$ gives
\begin{equation}\nonumber
\begin{split}
  dI=&\oint\Bigg\{\left[\frac{dk}{du}\delta x-\left(\frac{\prt\om}{\prt k}\frac{dk}{du}
  +\frac{\prt\om}{\prt u}\right)\delta t\right]\\
  &\times\left(\frac{\prt u}{\prt t}dt+\frac{\prt u}{\prt x}dx \right)-
  \delta x\cdot dx+\delta\om\cdot dt\Bigg\}=0,
  \end{split}
\end{equation}
where we have integrated the terms $kd\delta x-\om d\delta t$ by parts over the closed contour.
Now we substitute $dx=V_0\chi d\mu$, $dt=\chi d\mu$ and obtain after simple transformations with
account of Eq.~(\ref{eq16}) the expression
\begin{equation}\nonumber
  \Bigg\{\oint\left[\left(\frac{\prt\om}{\prt k}-V_0\right)\frac{dk}{du}+\frac{\prt\om}{\prt u}\right]
  \left(\frac{\prt u}{\prt t}\delta t+\frac{\prt u}{\prt x}\delta x \right)\chi\Bigg\}\cdot d\mu=0.
\end{equation}
This integral must vanish for any choice of $\chi$, so we get the equation
\begin{equation}\label{eq30}
  \frac{dk}{du}=\frac{\prt\om/\prt u}{V_0(u)-\prt\om/\prt k}.
\end{equation}
This equation was obtained earlier from analysis of Whitham equations at the small-amplitude edge
of DSWs \cite{el-05} and from Hamilton equations for propagation of wave packets along a simple wave
background pulse \cite{kamch-20a,ks-21}. Presented here derivation gives important new information
that the background evolution preserves the value of the Poincar\'e-Cartan integral (\ref{eq25}).

Let us assume now that we have found such a solution
\begin{equation}\label{eq31}
  k=k(u,\oq),
\end{equation}
($\oq$ being an integration constant) of Eq.~(\ref{eq30}) that the closed contour $C$ corresponds to
the path of the small-amplitude edge gone first in one direction with negative $k$ and then in the
opposite direction with positive $k$ according to the Hamilton
equations (\ref{eq17}). Then we can define ``number of oscillations''
\begin{equation}\label{eq32}
  N(\oq)=\frac{1}{4\pi}\oint_{C(\oq)}(k\delta x-\om\delta t)
\end{equation}
in the DSW corresponding to this value $\oq$ (additional factor $1/2$ compared with Eq.~(\ref{eq4}) 
is introduced for taking into
account that in the case of a closed contour the edge goes its path twice). This number $N(\oq)$
remains the same, when we transform the contour to the initial state at $t=0$ ($\delta t=0$):
\begin{equation}\label{eq32b}
  N(\oq)=\frac{1}{4\pi}\oint_{C_0(\oq)}k\delta x=\frac{1}{2\pi}\int_{x_1(\oq)}^{x_2(\oq)}k(u_0(x),\oq)dx,
\end{equation}
where $x_{1,2}(\oq)$ are two `turning point' at which
\begin{equation}\label{eq33}
  k(u_0(x_{1,2}(\oq)),\oq)=0
\end{equation}
and $u_0(x)$ is the initial distribution of the background pulse. The asymptotic expression
(\ref{eq8}) for the number of solitons assumes that these turning point go to infinities, but
less formal consideration makes it clear at once that there exists the maximal integer value of
$N$ for which the expression (\ref{eq32b}) with a given distribution $u_0(x)$ still has the
corresponding `eigenvalue' $\oq_N$ but for $N+1$ such an eigenvalue does not exist anymore.
In this case the integration interval $(x_1(\oq_N),x_2(\oq_N))$ is so wide for intensive initial
pulses with $u_0(x)\to0$ as $|x|\to\infty$, that it can be replaced with good enough accuracy by
the interval $(-\infty,\infty)$. We can say that $\oq_N$ corresponds to the $N$-th soliton in
the asymptotic state. If we decrease $N$ by one, then we get $N(\oq_{N-1})=N-1$, that is $\oq_{N-1}$
corresponds to $(N-1)$-th soliton, and so on. As a result, we arrange correspondence between
the $n$-th soliton and the parameter $\oq_n$ which is to be determined from the Bohr-Sommerfeld
quantization rule
\begin{equation}\label{eq34}
  \int_{x_1(\oq_n)}^{x_2(\oq_n)}k(u_0(x),\oq_n)dx=2\pi n,\qquad n=1,2,\ldots,N,
\end{equation}
where $k=k(u,\oq)$ is the solution (\ref{eq31}) of Eq.~(\ref{eq30}). At last, we notice that
Eq.~(\ref{eq34}) can be interpreted as a quasiclassical quantization rule for eigenvalues
of a linear wave equation for waves propagating along the ``potential'' $u_0(x)$. As is well
known from quantum mechanics (see, e.g., \cite{LL-3}), a more accurate description of wave
solutions in vicinities of the turning points leads to the replacement $n\mapsto n-1/2$,
and this is very general phenomenon of the short wavelength asymptotic behavior at $k\to\infty$
(see, e.g., Appendix~11 in Ref.~\cite{arnold-89}). We will make such a replacement in practical
applications of this theory without derivation just as a heuristic approximation confirmed by
comparison with numerical results.

We will relate the parameters $\oq_n$ obtained from the Bohr-Sommerfeld rule (\ref{eq34})
with the physical parameters of the asymptotic solitons in the next Section and now we turn to
generalization of this quantization rule to not simple-wave background flows.

\subsection{General flow of background pulse}

Now we suppose that the dispersionless flow is described by the solution $\rho=\rho(x,t)$,
$u=u(x,t)$ of Eqs.~(\ref{eq10}) with some initial distributions $\rho=\rho_0(x), u=u_0(x)$.
The dispersion relation in the Hamilton equations (\ref{eq17}) for the packet's propagation
depends on the values $\rho$ and $u$ at the point $x$ of the packet's instant location at the
moment $t$, so the function
\begin{equation}\label{eq36}
  \om=\om(k,\rho,u)
\end{equation}
is known from the linearized equations. We look again for such a function $k=k(\rho,u)$ that the
Poincar\'e-Cartan integral (\ref{eq25}) is preserved by the flow in the extended phase space
$(x,k,t)$ provided the contour $C$ is taken around a tube of streamlines defined by the equation
$dx/dt=u(x,t)$. As in the simple wave case of the background flow, we define a coordinate $\mu$
along streamlines by the equations
\begin{equation}\label{eq37}
  \frac{dx}{u}=dt=\chi d\mu,
\end{equation}
where $\chi=\chi(x,t,k)$ is an arbitrary function, and we demand the integral does not depend
neither on $\chi$ (that is the choice of the contour's form), no on $\mu$ (that is on position of
this contour on the tube). Differentiation of Eq.~(\ref{eq25}) with respect to $\mu$ gives with
account of Eqs.~(\ref{eq10}) and (\ref{eq17}) the expression
\begin{equation}\nonumber
  \begin{split}
  dI=\oint &\Bigg\{\left[\frac{\prt k}{\prt\rho}\left(\frac{\prt\om}{\prt k}-u\right)-
  \frac{c^2}{\rho}\frac{\prt k}{\prt u}+\frac{\prt\om}{\prt\rho}\right]\rho_x\\
  &+\left[\frac{\prt k}{\prt u}\left(\frac{\prt\om}{\prt k}-u\right)-
  {\rho}\frac{\prt k}{\prt\rho}+\frac{\prt\om}{\prt u}\right]u_x\Bigg\}\\
  &\times(\delta x-u\delta t)\chi\cdot d\mu=0.
  \end{split}
\end{equation}
Since $\chi$ is an arbitrary function, the expression in curly brackets must vanish. Moreover,
the expressions in square brackets are only functions of $\rho$ and $u$, whereas the local
values of $\rho_x$ and $u_x$ depend on the choice of the initial distributions, so they can be
considered as arbitrary functions, too. Hence, the expressions in square brackets vanish
separately and we arrive at the equations
\begin{equation}\label{eq38}
\begin{split}
&\frac{\prt k}{\prt \rho} = \frac{(v_g-u)\frac{\prt \omega}{\prt \rho} +
\frac{c^2}{\rho}\frac{\prt \omega}{\prt  u}}{c^2-(v_g - u)^2},\\
&\frac{\prt k}{\prt u} = \frac{(v_g-u)\frac{\prt \omega}{\prt u} +
\rho\frac{\prt \omega}{\prt  \rho}}{c^2-(v_g - u)^2},
\end{split}
\end{equation}
which should determine the function $k=k(\rho,u)$ for given dispersion relation (\ref{eq36})
and equation of state $c=c(\rho)$. Eqs.~(\ref{eq38}) have recently been derived in
Ref.~\cite{sk-23} from the Hamilton equations (\ref{eq17}) under the same supposition that $k$
is only a function of $\rho$ and $u$.

For existence of the function $k=k(\rho,u)$, its derivatives defined by Eqs.~(\ref{eq38})
must satisfy the compatibility condition
\begin{equation}\label{eq39}
  \frac{\prt}{\prt \rho}\left(\frac{\prt k}{\prt u}\right)=
  \frac{\prt}{\prt u}\left(\frac{\prt k}{\prt \rho}\right).
\end{equation}
If this condition is not fulfilled identically, we can confine ourselves to the limit of large
values of $k$ since just this limit corresponds to the high quantum numbers in the
Bohr-Sommerfeld quantization rule. To this end, we expand the right-hand sides of Eqs.~(\ref{eq38})
with respect to a small parameter $\sim c/k$ and obtain
\begin{equation}\label{eq40}
\begin{split}
  \frac{\prt k}{\prt \rho}=R_{1}(k,\rho,u),\qquad
  \frac{\prt k}{\prt u}=R_{2}(k,\rho,u),
  \end{split}
\end{equation}
where in $R_{1,2}$ only the terms are held that satisfy the condition (\ref{eq39}). Then these
equations allow one to restore the function
\begin{equation}\label{eq41}
  k=k(\rho,u,q),
\end{equation}
where $q$ is an integration constant determined by the initial value $k=k_0\gg c$. It is
convenient to define $q$ in such a way that the condition $k\sim k_0\gg c$ corresponds to $q\gg c$.

The solution (\ref{eq41}) (exact of approximate) allows one to define the Poincar\'e-Cartan
integral
\begin{equation}\label{eq42}
  N(q)=\frac{1}{4\pi}\oint_{C}(k\delta x-\om\delta t)
\end{equation}
for contours $C$ around a tube of streamlines of the dispersionless flow. If we transform such a
contour to the initial state at $t=0$, we obtain
\begin{equation}\label{eq43}
  N(q)=\frac{1}{4\pi}\oint_{C_0(q)}k\delta x=\frac{1}{2\pi}\int_{x_1(q)}^{x_2(q)}k[\rho_0(x),u_0(x),q]dx,
\end{equation}
where $x_{1,2}(q)$ are the turning points defined by the equation
\begin{equation}\label{eq44}
  k[\rho_0(x_{1,2}(q),u_0(x_{1,2}(q)),q]=0.
\end{equation}
If we only know the asymptotic solution $k\gg c$ of Eq.~(\ref{eq40}), then Eq.~(\ref{eq44})
give only approximate solution for the turning points. Nevertheless, Eq.~(\ref{eq43}) gives
accurate enough value of the integral, since along the most part of the integration interval
we have $k\gg c$ and vicinities of the turning points give negligibly small contribution to
the integral. As in the simple-wave background pulse case, we arrange correspondence between
the soliton's number $n$ and the value $q_n$ determined by the Bohr-Sommerfeld rule
\begin{equation}\label{eq45}
  \int_{x_1(q_n)}^{x_2(q_n)}k[\rho_0(x),u_0(x),q_n)dx=2\pi n,\quad n=1,2,\ldots,N.
\end{equation}
In practical applications of this formula we will make a replacement $n\mapsto n+1/2$,
as it was explained below Eq.~(\ref{eq34}).

Now we have to relate the constants $\oq_n$ (simple-wave case) or $q_n$ (general case) with
parameters of asymptotic solitons emerging eventually from a given initial pulse.

\section{Parameters of asymptotic solitons}

Now, when each soliton is labeled by a specific value of the parameter $\oq_n$ or $q_n$, we need to
relate it with soliton's physical parameters, as, for example, its velocity $V_n$. This can be done
with the use of Stokes remark \cite{lamb,stokes} (see also \cite{ai-77,dkn-03}) that the small amplitude
tails $\propto\exp[\pm\tk(x-Vt)]$ of a soliton obey the same linearized equations as a small-amplitude
harmonic wave $\propto\exp[i(kx-\om t)]$. Consequently, the soliton's velocity $V_n$ can be
expressed in the form
\begin{equation}\label{eq46}
  V_n=\tom(\tk_n)/\tk_n,\quad\text{where}\quad \tom(\tk)=i\om(-i\tk),
\end{equation}
and $\tk$ denotes the soliton's inverse half-width (see Refs.~\cite{el-05,kamch-20a,kamch-21}).
This means analytical continuation of the function $\om=F(k^2)$ from the interval of positive
values of its argument $k^2$ to the interval of its negative values of the argument $k^2=-\tk^2<0$.
In practice it means that we
consider the same function $F$ at different intervals of its argument $k^2$.

We generalize this Stokes observation to other functions of $k^2$ that can also be continued to
the region $k^2<0$ where they express some assertions about the inverse half-widths $\tk$ of
solitons. For example, suppose we have found the exact solution (\ref{eq31}) of Eq.~(\ref{eq30})
and it can be expressed in the form
\begin{equation}\label{eq47}
  k^2=\overline{K}(u,\oq).
\end{equation}
%In Refs.~\cite{el-05,kamch-20a,kamch-19} some reasons were presented that
This formula can be
continued to the solitonic region to give
\begin{equation}\label{eq48}
  \tk^2=-\overline{K}(u,\oq).
\end{equation}
Suppose we have found from the Bohr-Sommerfeld rule the value $\oq_n$ for the $n$-th soliton and
at asymptotically large time this soliton propagates along zero background $u=0$. Then its
inverse half-width is given by the formula
\begin{equation}\label{eq49}
  \tk_n^2=-\overline{K}(0,\oq_n)
\end{equation}
and, consequently, its velocity is readily obtained from Eq.~(\ref{eq46}). This approach
reproduces in a different form the theory developed for simple wave pulses
in Refs.~\cite{egs-08,mfweh-20}.

We can extend the above approach to the case of exact solutions of Eqs.~(\ref{eq38}),
so we imply that the condition (\ref{eq39}) is fulfilled. Again we express the exact solution
in the form
\begin{equation}\label{eq50}
  k^2=K(\rho,u,q)
\end{equation}
and analytically continue it to the soliton region to obtain
\begin{equation}\label{eq51}
  \tk^2=-K(\rho,u,q).
\end{equation}
If we found for the $n$-th soliton the eigenvalue $q_n$ from the Bohr-Sommerfeld rule
(\ref{eq45}) and if the asymptotic soliton propagates along the uniform background with
$\rho=\rho_R,u=0$, then its inverse half-width is given by the formula
\begin{equation}\label{eq52}
  \tk_n^2=-K(\rho_R,0,q_n)
\end{equation}
and its velocity is to be found from the Stokes rule (\ref{eq46}).

However such an analytical continuation becomes impossible in case of asymptotic solutions
(\ref{eq41}) which are only correct for $k\gg c\sim u$. Nevertheless, we can overcome this
difficulty in the
following way. Suppose we have found the asymptotic solution in the form (\ref{eq50}). Naturally,
this general solution remains correct in case of simple-wave pulses when there is the
functional relationship between $\rho$ and $u$, say, $\rho=\rho_{\mathrm{sw}}(u)$, so we get the
asymptotic formula
\begin{equation}\label{eq53}
  k^2=K(\rho_{\mathrm{sw}}(u),u,q),\qquad k\gg u.
\end{equation}
Now, the asymptotic soliton trains propagate along simple-wave solutions, so their inverse
half-widths are asymptotic specifications of the formula (\ref{eq48}) obtained by analytic
continuation of Eq.~(\ref{eq47}). If we take its series expansion with respect to
small parameter $u$, then the resulting formula
\begin{equation}\label{eq54}
  k^2=\overline{K}_{\text{asymp}}(u,\oq)
\end{equation}
must coincide with Eq.~(\ref{eq53}). This matching condition allows one to find the relationship
between the parameters $q$ and $\oq$:
\begin{equation}\label{eq55}
  \oq=\oq(q).
\end{equation}
Consequently, the inverse half-widths of asymptotic solitons are given by Eq.~(\ref{eq49}),
\begin{equation}\label{eq56}
  \tk_n^2=-\overline{K}(0,\oq(q_n)),
\end{equation}
where the parameters $q_n$ are to be obtained from the Bohr-Sommerfeld rule (\ref{eq45}) and we
assumed that the asymptotic solitons propagate along the background with $u=0$. At last,
the soliton's velocities are to be found according to the Stokes rule (\ref{eq46}).

Let us illustrate this theory by examples.

\section{Solitons evolved from a large deep in the generalized NLS equation theory}

Here we shall consider evolution of a pulse according to the defocusing generalized NLS
equation (generalized Gross-Pitaevskii equation for a Bose-Einstein condensate of atoms
with repulsive interaction)
\begin{equation}\label{eq57}
i\psi_t+\frac{1}{2}\psi_{xx} - f(|\psi|^2)\psi = 0,
\end{equation}
where the nonlinearity function is positive: $f(\rho)>0$, $f(0)=0$. It is well known
that by means of the substitution 
\begin{equation}\label{eq57a}
  \psi=\sqrt{\rho}\exp\left(i\int^xu(x',t)dx'\right)
\end{equation}
this equation can be transformed to the hydrodynamic-like system
\begin{equation}\label{eq58}
\begin{split}
&\rho_t + (\rho u)_x = 0, \\
&u_t + uu_x + \frac{c^2}{\rho} \rho_x + \bigg( \frac{\rho_x^2}{8\rho^2}
- \frac{\rho_{xx}}{4\rho} \bigg)_x = 0,
\end{split}
\end{equation}
where
\begin{equation}\label{eq59}
  c^2=\rho f'(\rho).
\end{equation}
Linearization of this system with respect to small deviations from a uniform flow
yields the Bogoliubov dispersion relation
\begin{equation}\label{eq60}
\omega = k\left( u \pm \sqrt{c^2+\frac{k^2}{4}} \right),
\end{equation}
that is $c$ has the meaning of the sound velocity in the dispersionless limit $k\to0$.
Equations (\ref{eq58}) reduce in the same limit to the standard Euler equations
(\ref{eq10}).

Now we can check whether the derivatives (\ref{eq38}) commute.
In this case it is convenient to replace $\rho$ by the variable $c$ where the function
$\rho=\rho(c)$ is obtained by inversion of the function $c=c(\rho)$ (see Eq.~(\ref{eq59}))
so Eqs.~(\ref{eq38}) are cast to the form
\begin{equation}\label{eq61}
  \begin{split}
  & \frac{\prt k}{\prt c}=-\frac{c[(2+c\rho'/\rho)k^2+4(1+c\rho'/\rho)c^2]}{k(k^2+3c^2)},\\
  & \frac{\prt k}{\prt u}=-\frac{\sqrt{k^2+4c^2}[k^2+2(1+\rho/(c\rho'))c^2]}{k(k^2+3c^2)}.
  \end{split}
\end{equation}
The straightforward calculation yields (see also \cite{sk-23})
\begin{equation}\label{eq62}
  \begin{split}
  &\frac{\prt}{\prt c}\left(\frac{\prt k}{\prt u}\right)-\frac{\prt}{\prt u}\left(\frac{\prt k}{\prt c}\right)
  =\frac{\sqrt{k^2+4c^2}}{k\rho\rho'(k^2+3c^2)^2}\\
  &\times\left[(k^2+6c^2)\rho^{\prime 2}(\rho'-2c)+2(k^2+3c^2)\rho^2(c\rho^{\prime\prime}-\rho')\right],
  \end{split}
\end{equation}
and we see that the compatibility condition is only fulfilled for the case of the Kerr-like
nonlinearity with $f(\rho)=\rho$ and $\rho=c^2$. Otherwise, it is fulfilled only in the limit
$k\to\infty$, so it is natural to consider these two situations separately.

\subsection{Kerr-like nonlinearity $f(\rho)=\rho$}

NLS equation with Kerr-like nonlinearity is completely integrable \cite{zs-73}, so the Bohr-Sommerfeld
quantization rule for solitons parameters can be derived from the Zakharov-Shabat linear spectral
problem \cite{JLML-99,kku-02}. It is instructive to see, how these known results follow from our
general approach not based on the complete integrability property. We assume here that the initial pulse
is represented by a deep in the density or sound velocity distribution $c=c_{0}(x)$ and there is also
some distribution of the initial flow velocity $u=u_{0}(x)$. Far enough from the initial pulse
we have $c_{0}(x)\to c_R$, $u_{0}(x)\to0$ as $|x|\to\infty$. Naturally, these two distributions
are equivalent to some initial distributions $r_{\pm}^{(0)}(x)$ of the dispersionless Riemann invariants
(see (\ref{eq14}))
\begin{equation}\label{eq63}
  r_{\pm}=\frac{u}{2}\pm c.
\end{equation}

In this case with $\rho=c^2$ Eqs.~(\ref{eq61}) can be written in the form
\begin{equation}\label{eq64}
\begin{split}
\frac{\prt k}{\prt c} = -4\frac{c}{k},\qquad
\frac{\prt k}{\prt u} = -\frac{\sqrt{k^2+4c^2}}{k},
\end{split}
\end{equation}
and they have the exact solution
\begin{equation}\label{eq65}
  k^2=(q-u)^2-4c^2,
\end{equation}
where $q$ is an integration constant. If we write it in the form
\begin{equation}\label{eq66}
k^2=4\left(\frac{q}{2}-\frac{u}{2}-c\right)\left(\frac{q}{2}-\frac{u}{2}+c\right)=
4(\la-r_+)(\la-r_-),
\end{equation}
where we have defined $\la=q/2$, then the Bohr-Sommerfeld rule (\ref{eq45})
reads
\begin{equation}\label{eq67}
\begin{split}
  \int_{x_1(\la_n)}^{x_2(\la_n)}
  \sqrt{(\la_n-r_+^{(0)}(x))(\la_n-r_-^{(0)}(x))}\,dx=n\pi,\\
   n=1,2,\dots, N.
  \end{split}
\end{equation}
It coincides in the main approximation in the limit $n\to\infty$ with the asymptotic
formula for the linear spectral problem eigenvalues (see \cite{JLML-99,kku-02}).
It is worth, however, notice that in better approximation developed in Ref.~\cite{JLML-99}
the integer $n$ should be replaced by $n+1/2$ with $n=1,2,\ldots,N$ in agreement with
the replacement $n\mapsto n+1/2$ mentioned below Eq.~(\ref{eq34}).

Since Eq.~(\ref{eq65}) represents the exact solution of Eqs.~(\ref{eq64}), it is
correct for any values of $k$ and can be continued to the soliton region, so we obtain
\begin{equation}\label{eq68}
  \tk^2=-4(\la-r_+)(\la-r_-).
\end{equation}
In the asymptotic region, when solitons propagate along background with $r_{\pm}=\pm c_R$,
we get for the half-width of the $n$-th soliton the expression
$\tk_n^2=4(c_R^2-\la_n^2)$. Consequently, according to the Stokes rule (\ref{eq46}),
its velocity is equal to
\begin{equation}\label{eq69}
  V_n=\frac{\tom(\tk_n)}{\tk_n}=\sqrt{c_R^2-\frac{\tk^2}{4}}=\la_n
\end{equation}
in agreement with the known results (see, e.g., Ref.~\cite{kku-02}).

The possibility to find the exact solution of Eqs.~(\ref{eq61}) in this case is related,
apparently, with complete integrability of the NLS equation (\ref{eq57}) with Kerr
nonlinearity $f(\rho)=\rho$. We will clarify this point in Section \ref{integrability}.

\subsection{Non-Kerr nonlinearity}

In the limit of large $k\gg c$ the right-hand side of Eq.~(\ref{eq62}) tends to zero as
$\propto k^{-2}$, that is the condition (\ref{eq39}) is fulfilled in this limit and we can
find the asymptotic expression for the function $k=k(c,u)$. Eqs.~(\ref{eq61}) in the limit
of large $k$ can be written as
\begin{equation}\label{eq70}
  \frac{\prt k^2}{\prt c^2}=-2\left(1+\frac{c^2}{\rho}\frac{d\rho}{d c^2}\right),\qquad
  \frac{\prt k}{\prt u}=-1.
\end{equation}
The second equation gives $k=q-u+F(c^2)$, where in the main approximation
$k\approx q\sim k_0\gg|u|,c$. We suppose that $F(c^2)\sim c^2$, so $k^2\approx(q-u)^2+2qF(c^2)$,
where we have neglected small terms $uF\sim c^3$ and $F^2\sim c^4$. Then the first equation
(\ref{eq70}) gives
$$
F(c^2)=-\frac1q\left(c^2+\int_0^{\rho(c)}\frac{c^2}{\rho}d\rho\right)\sim\frac{c^2}{q}
$$
in agreement with our supposition about the order of magnitude of $F$. Thus, we obtain
the asymptotic solution
\begin{equation}\label{eq71}
  k^2=(q-u)^2-2\left(c^2+\int_0^{\rho(c)}\frac{c^2}{\rho}d\rho\right).
\end{equation}
For a particular case of the nonlinearity function
\begin{equation}\label{eq72}
  f(\rho)=\frac1{\ga-1}\rho^{\ga-1},\quad\text{when}\quad c^2=\rho^{\ga-1},\quad \rho=c^{2/(\ga-1)},
\end{equation}
we get
\begin{equation}\label{eq73}
  k^2=(q-u)^2-\frac{2\ga}{\ga-1}c^2.
\end{equation}
If $\ga=2$, that is $f(\rho)=\rho$, this asymptotic formula reproduces the exact solution
(\ref{eq65}). The function $k=k(c,u)$ defined by Eq.~(\ref{eq73}) can be used in the Bohr-Sommerfeld
rule (\ref{eq45}), so we get
\begin{equation}\label{eq74}
\begin{split}
  &\int_{x_1(q_n)}^{x_2(q_n)}\sqrt{(q_n-u_0(x))^2-\frac{2\ga}{\ga-1}c_0^2(x)}\,\,dx\\
  &=2\pi\left(n+\frac12\right),\qquad   n=1,2,\ldots,N,
   \end{split}
\end{equation}
where %$c=c_0(x),u=u_0(x)$ are the initial distributions of $c$ and $u$, and
we have made the replacement
$n\mapsto n+1/2$. As a result, we obtain a set of parameters $q_n$ for
solitons emerging from the pulse with given initial distributions $c=c_0(x), u=u_0(x)$.
To relate $q_n$ with velocities of these solitons, we should turn to the simple-wave solution
of the corresponding equation (\ref{eq30}).

The problem of evolution of a step-like discontinuity of the simple-wave type was studied in much
detail in Ref.~\cite{hoefer-14} by the method of Ref.~\cite{el-05}. The solution of
Eq.~(\ref{eq30}) found there can be written in the form
\begin{equation}\label{eq75}
  k^2=4c^2(\al^2(c,\oq)-1),
\end{equation}
where the function $\al=\al(c,\oq)$ is defined in implicit form in our notation by the formula
\begin{equation}\label{eq76}
  \frac{c}{\oq}=\left(\frac{2}{1+\al}\right)^{\frac{\ga-1}{3\ga-5}}
  \left(\frac{\ga+1}{3-\ga+2(\ga-1)\al}\right)^{\frac{2(\ga-2)}{3\ga-5}}.
\end{equation}
In the limit $c\ll\oq$ we have $\al\gg 1$, so we obtain the series expansion
\begin{equation}\label{eq77}
  \frac{c}{\oq}=\beta(\ga)\left(\frac{1}{\al}-\frac{1}{\ga-1}\cdot\frac{1}{\al^2}+
  \frac{\ga^2-3\ga+6}{4(\ga-1)^2}\cdot\frac{1}{\al^3}+\ldots\right),
\end{equation}
where
\begin{equation}\label{eq78}
  \beta(\ga)=2^{\frac{3-\ga}{3\ga-5}}\left(\frac{\ga+1}{\ga-1}\right)^{\frac{2(\ga-2)}{3\ga-5}}.
\end{equation}
Inversion of this series and substitution of the result into Eq.~(\ref{eq75}) yield with
necessary accuracy
\begin{equation}\label{eq79}
\begin{split}
  k^2\approx &(2\oq\beta)^2-\frac{4}{\ga-1}(2\oq\beta)c\\
  &+\left(\frac{4}{(\ga-1)^2}-\frac{2}{\ga-1}-2\right)c^2,\qquad c\ll\oq.
  \end{split}
\end{equation}
This asymptotic expression should be matched with Eq.~(\ref{eq73}) for the simple wave pulses.

The dispersionless Riemann invariants (\ref{eq14}) in our case (\ref{eq72}) are equal to
\begin{equation}\label{eq80}
  r_{\pm}=\frac{u}{2}\pm\frac{c}{\ga-1}.
\end{equation}
We consider the asymptotic train of right-propagating solitons along the background with
$$
r_-=\frac{u}{2}-\frac{c}{\ga-1}=-\frac{c_R}{\ga-1}=\mathrm{const},
$$
where $c\to c_R$, $u\to0$ as $|x|\to\infty$. Hence, we have to substitute
\begin{equation}\label{eq81}
  u=\frac{2}{\ga-1}(c-c_R)
\end{equation}
into Eq.~(\ref{eq73}) to obtain
\begin{equation}\label{eq82}
\begin{split}
  k^2=&\left(q+\frac{2c_R}{\ga-1}\right)^2-\frac{4}{\ga-1}\left(q+\frac{2c_R}{\ga-1}\right)c\\
  &+\left(\frac{4}{(\ga-1)^2}-\frac{2}{\ga-1}-2\right)c^2,\qquad c\ll q.
  \end{split}
\end{equation}
This expression coincides with Eq.~(\ref{eq79}) for
\begin{equation}\label{eq83}
  \oq=\frac{1}{2\beta}\left(q+\frac{2c_R}{\ga-1}\right).
\end{equation}
In the solitonic region Eq.~(\ref{eq75}) gives
\begin{equation}\label{eq84}
  \tk^2=4c^2(1-\al^2(c,\oq)),
\end{equation}
so the asymptotic velocities of solitons propagating along a uniform background with $c=c_R$,
$u=0$ are equal to
\begin{equation}\label{eq85}
\begin{split}
  V_n&=\frac{\tom(\tk_n)}{\tk_n}=\sqrt{c_R^2-\frac{\tk_n^2}{4}}\\
  &=c_R\al\left(c_R,\frac{1}{2\beta}\left(q_n+\frac{2c_R}{\ga-1}\right)\right),
  \end{split}
\end{equation}
where $q_n$ are to be obtained from the Bohr-Sommerfeld quantization rule (\ref{eq74}).

Soliton's velocity $V$ is related with minimal value $\rho_m$ of the density at its center by 
the formula (see Ref.~\cite{ks-09})
\begin{equation}\label{eq85b}
  V^2=\frac{2\rho_m}{(\rho_R-\rho_m)^2}\int_{\rho_m}^{\rho_R}\left[f(\rho_R)-f(\rho)\right]d\rho,
\end{equation}
so when the soliton's velocity is known, we can easily find the amplitude of this soliton.

Let us illustrate the theory by a concrete example $\ga=3$, when $\beta=\sqrt{2}$ and the function
$\al=\al(c)$ can be found in the explicit form (see Ref.~\cite{hoefer-14}),
\begin{equation}\label{eq86}
  \al(c,\oq)=\frac12\left(\sqrt{1+8\left(\frac{\oq}{c}\right)^2}-1\right).
\end{equation}
As a result, we obtain a simple formula
\begin{equation}\label{eq87}
  V_n=\frac12\left(\sqrt{c_R^2+(c_R+q_n)^2}-c_R\right).
\end{equation}

\begin{figure}[t]
\begin{center}
	\includegraphics[width = 8cm,height = 6cm]{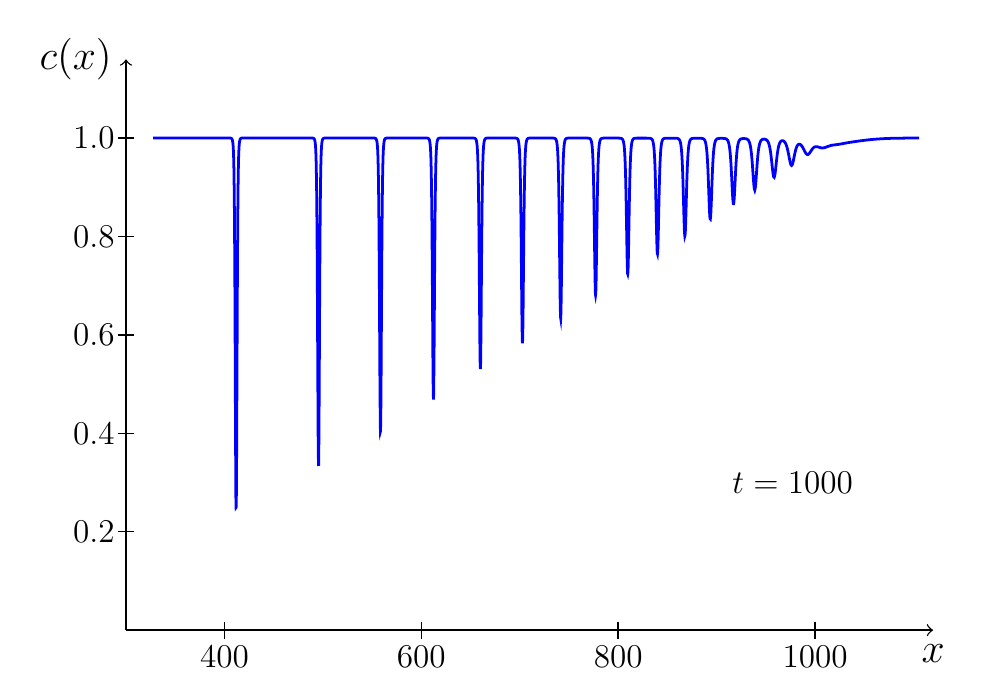}
\caption{Distribution of $c(x)$ in the right-propagating soliton train evolved from
the pulse with the initial distributions (\ref{eq88}) at $t=1000$.
 }
\label{fig1}
\end{center}
\end{figure}

We compared these analytical predictions with results of the exact numerical solution of the
gNLS equation (\ref{eq57}) with $f(\rho)=\rho^2/2$ ($\ga=3$) for the initial distributions
\begin{equation}\label{eq88}
  c_0(x)=1-\frac{0.9}{\cosh^2(x/20)},\qquad u_0(x)=0.
\end{equation}
In this case the initial pulse evolves into two symmetrical right- and left-propagating dark
soliton trains and a typical distribution of $c(x)$ at $t=1000$ is shown in Fig.~{\ref{fig1}
for the right-propagating solitons. If we assume that all the solitons started their motion
at $x=0$ at the moment $t=0$ with asymptotic velocities $V_n$, then their coordinates at the
moment $t$ are equal to $x_n(t)=t(\sqrt{1+(1+q_n)^2}-1)/2$. This formula allows one to find
$q_n$ from numerical values of the coordinates,
\begin{equation}\label{eq89}
  q_n^{\text{num}}=\sqrt{\left(2x_n(t)/t+1\right)^2-1}-1.
\end{equation}
These values can be compared with the values $q_n^{\text{B.-S.}}$ found from the Bohr-Sommerfeld
rule (\ref{eq74}) in a particular case of the initial distributions (\ref{eq88}), and the
results are shown in Fig.~\ref{fig2}, where crosses correspond to $q_n^{\text{B.-S.}}$ and
dots to $q_n^{\text{num}}$. As one can see, the agreement is quite good even for small values of $n$.

\begin{figure}[t]
\begin{center}
	\includegraphics[width = 8cm,height = 6cm]{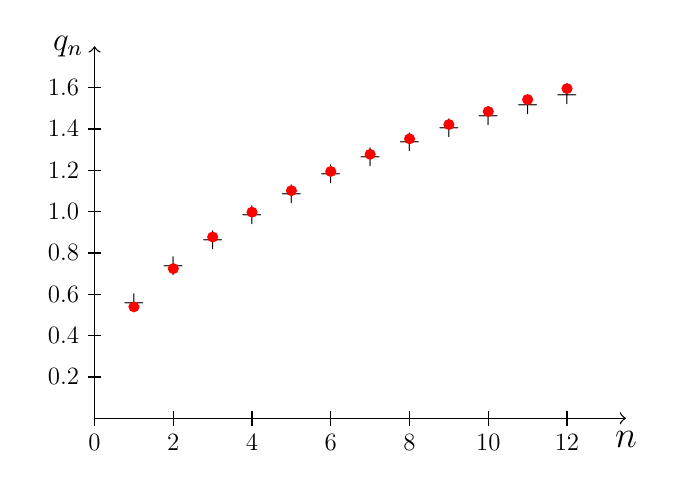}
\caption{Parameters $q_n^{\text{B.-S.}}$ obtained from the Bohr-Sommerfeld rule (crosses)
and $q_n^{\text{num}}$ obtained from the numerical solution (red dots).
 }
\label{fig2}
\end{center}
\end{figure}

In case of $f(\rho)=\rho^2/2$ Eq.~(\ref{eq85b}) gives the expression for the minimal density
at the $n$-th soliton
\begin{equation}\label{eq89b}
  \rho_m^{(n)}=\left[\rho_R^2+\frac34\left(\sqrt{\rho_R^2+(q_n+\rho_R)^2}-\rho_R\right)^2\right]^{1/2}-\rho_R.
\end{equation}
This formula also agrees very well with the minimal values of density at centers of solitons 
shown in Fig.~\ref{fig1}.

\section{Connection with the theory of completely integrable equations}\label{integrability}

As we saw in the preceding Section, there is sharp difference between the cases with $\ga=2$
and $\ga\neq2$: if $\ga=2$ the derivatives (\ref{eq61}) commute and we get the exact solution
(\ref{eq65}) of these equations, whereas if $\ga\neq 2$ we can only obtain the asymptotic solution 
(\ref{eq73}) correct in the limit $q\gg|u|,c$. It seems very plausible that such a difference
between two situations is related with the complete integrability of the NLS equation (\ref{eq57})
with $f(\rho)=\rho$, ($\ga=2$). Here we shall consider this relationship for the class of 
completely integrable equations which belong to the Ablowitz-Kaup-Newell-Segur (AKNS) 
scheme \cite{akns-74}.

Usually the AKNS scheme is formulated in $2\times2$-matrix form, but for discussion of its 
quasiclassical limit it is convenient to use its scalar form (see Ref.~\cite{ak-02}). So we assume 
that an integrable equation under consideration can be written as a compatibility condition of two
linear equations
\begin{equation}\label{eq90}
  \phi_{xx}=\mathcal{A}\phi,\qquad \phi_t=-\frac12  \mathcal{B}_x\phi+ \mathcal{B}\phi_x
\end{equation}
for the function $\phi$, where $\mathcal{A}$ and $\mathcal{B}$ depend on the wave variables
(let them be $\rho$ and $u$) and the spectral parameter $\la$. Then the compatibility condition
$(\phi_{xx})_t=(\phi_t)_{xx}$ leads to the equation
\begin{equation}\label{eq91}
\mathcal{A}_t-2\mathcal{B}_x\mathcal{A}-\mathcal{B}\mathcal{A}_x+
\frac12\mathcal{B}_{xxx}=0,
\end{equation}
which must be fulfilled for any values of $\la$, so we get the nonlinear equations under 
consideration. This provides a number of useful relations for solutions without actual
finding them. In particular, the first Eq.~(\ref{eq90}) has two basis solutions $\phi_+,\phi_-$,
so we define the function
\begin{equation}\label{eq92}
  g=\phi_+\phi_-,
\end{equation}
which satisfies the equations
\begin{equation}\label{eq93}
 g_{xxx}-2\mathcal{A}_x{g}-4\mathcal{A}{g}_x=0, \qquad
{g}_t=\mathcal{B}{g}_x-\mathcal{B}_x{g}.
\end{equation}
The first one is readily integrated to give
\begin{equation}\label{eq94}
\frac12{g}{g}_{xx}-\frac14{g}_x^2-\mathcal{A}{g}^2={P},
\end{equation}
where $P$ is an integration constant. Then the second Eq.~(\ref{eq93}) yields
\begin{equation}\label{eq95}
\left(\frac{\sqrt{P}}{{g}}\right)_t=
\left(\frac{\sqrt{P}}{{g}}\mathcal{B}\right)_x,
\end{equation}
where we have introduced the constant factor $\sqrt{P}$ under the differentiation signs to fix
the asymptotic behavior of $g$ in the limit $\la\to\infty$. Eq.~(\ref{eq95}) can serve as a
generating function of conservation laws of our nonlinear wave equations and averaging of
this generating function provides a convenient method of derivation of the Whitham modulation
equations \cite{kamch-94,kamch-04}. Solutions of the linear equations (\ref{eq90}) can also be
expressed in terms of the function $g$ (see, e.g., \cite{kku-02,ak-01})
\begin{equation}\label{eq96}
  \phi_{\pm}=\sqrt{g}\exp\left(\pm i\int^x\frac{\sqrt{P}}{g}\,dx\right).
\end{equation}
Now we can turn to the discussion of the quasiclassical limit of these equations.

Let us study transformation of a large-scale pulse to soliton trains. This transformation
can be represented as formation and evolution of dispersive shock waves (see, e.g., \cite{kamch-21a}).
According to Gurevich and Pitaevskii \cite{gp-73}, such an evolution can be represented as a slow 
change of modulation parameters in the periodic solution of equations under consideration what
leads to the slow variation of $P$ as well as parameters in $g$. Nevertheless, following to the
Whitham method \cite{whitham,Whitham-74} in Krichever's formulation \cite{krichever-88} (see also
Ref.~\cite{dn-93}), we assume that expressions (\ref{eq95}) and (\ref{eq96}) remain correct
during evolution of $\rho$ and $u$ from the initial smooth distributions to the asymptotic
soliton trains. When $\rho$ and $u$ are still smooth functions, the function $g$ is also
smooth, so the terms with derivatives in Eq.~(\ref{eq94}) can be neglected and we obtain
\begin{equation}\label{eq97}
  \frac{\sqrt{P}}{g}\approx\sqrt{-\overline{\mathcal{A}}}\equiv\overline{k}(\rho,u,\la),
\end{equation}
where $\overline{\mathcal{A}}=\overline{\mathcal{A}}(\rho,u)$ is to be obtained from $\mathcal{A}$
in the same approximation with omitted derivatives of $\rho$ and $u$. Then we arrive at a 
quasiclassical limit of Eq.~(\ref{eq96}),
\begin{equation}\label{eq98}
  \phi_{\pm}\approx\sqrt{g}\exp\left(\pm i\int^x\overline{k}(\rho,u,\la)\,dx\right).
\end{equation}
The condition that $\phi_{\pm}$ are single-valued functions of $x$ gives the Bohr-Sommerfeld
quantization rule (see, e.g., \cite{karpman-73,JLML-99,kku-02})
\begin{equation}\label{eq99}
\begin{split}
  \int_{x_1(\la_n)}^{x_2(\la_n)}\overline{k}(\rho_0(x),u_0(x),\la_n)\,dx=\pi\left(n+\frac12\right),\\
   n=1,2,\dots,
  \end{split}
\end{equation}
where ${x_1(\la_n)},{x_2(\la_n)}$ are the turning points. Comparison with Eq.~(\ref{eq45})
gives the expression for the wave number $k$ in our approach,
\begin{equation}\label{eq100}
  k(\rho,u)=2\overline{k}(\rho,u)=2\sqrt{-\overline{\mathcal{A}}(\rho,u)},
\end{equation}
as well as some relationship between the integration constant $q$ in our theory and the spectral
parameter $\la$ in the AKNS scheme. Consequently, in case of nonlinear wave equations, which are
completely integrable in the AKNS scheme, we obtain the expression for the function $k=k(\rho,u)$
without solving Eqs.~(\ref{eq38}).

Substitution of $\sqrt{P}/g=k/2$ into Eq.~(\ref{eq95}) gives the equation
\begin{equation}\label{eq101}
  k_t-(k\overline{\mathcal{B}})_x=0,
\end{equation}
where $\overline{\mathcal{B}}(\rho,u)$ is obtained from $\mathcal{B}$ in the same quasiclassical
approximation. This equation must coincide with the number of waves conservation law (\ref{eq2}),
so we get
\begin{equation}\label{eq102}
  \frac{\om}{k}=-\overline{\mathcal{B}}.
\end{equation}
As we see, the quasiclassical limits of the functions $\mathcal{A}$ and $\mathcal{B}$ are related
with the dispersion relation $\om=\om(k,\rho,u)$ of linear waves and the function $k=k(\rho,u,q)$
which can be obtained by solving Eqs.~(\ref{eq38}) if these derivatives commute. From this point
of view, the condition of commutativity of derivatives in Eqs.~(\ref{eq38}) can be considered as
an ``integrability test'' of nonlinear equations in framework of the AKNS scheme. If this
test is fulfilled, then equations 
\begin{equation}\label{eq103}
\begin{split}
  \overline{\mathcal{A}}(\rho,u,q)&=-\frac14k^2(\rho,u,q),\\ 
  \overline{\mathcal{B}}(\rho,u,q)&=-\frac{\om(k(\rho,u,q),\rho,u)}{k(\rho,u,q)}
  \end{split}
\end{equation}
give us the quasiclassical limit of the functions $\mathcal{A}$, $\mathcal{B}$ in the 
Lax pair (\ref{eq90}) and $q$ plays the role of the spectral parameter.

Let us apply the above consideration to the NLS equation (\ref{eq57}) with $f(\rho)=\rho$.
In this case we have (see Ref.~\cite{alber-93})
\begin{equation}\label{eq105}
\begin{split}
  \mathcal{A}&=-\left(\la+\frac{i\psi_x}{2\psi}\right)^2+\psi^*\psi-
  \left(\frac{\psi_{x}}{2\psi}\right)_x, \\ 
  \mathcal{B}&=-\la+\frac{i}2\frac{\psi_x}{\psi}.
  \end{split}
\end{equation}
The substitution of Eq.~(\ref{eq57a}) and neglecting derivatives of $\rho$ and $u$ yield
\begin{equation}\label{eq106}
\overline{\mathcal{A}}=-\left(\la-\frac{u}{2}\right)^2+\rho,\qquad 
\overline{\mathcal{B}}=-\la-\frac{u}{2}.
\end{equation}
Then Eq.~(\ref{eq100}) gives $k^2=4\left[(\la-u/2)^2-\rho\right]$ in agreement with 
Eq.~(\ref{eq65}) for $q=2\la$. Exclusion of $\la$ from this equation and from 
Eq.~(\ref{eq102}), that is from $\om=k(\la+u/2)$, reproduces the dispersion relation 
(\ref{eq60}).

\section{Conclusion}

The asymptotic method developed in this paper provides a simple way to finding parameters of
solitons evolved from given initial distributions of wave variables for a wide class of
nonintegrable nonlinear equations. Besides that, it presents 
a simple method of derivation of the asymptotic Bohr-Sommerfeld generalized rule for a linear 
spectral problem in framework of the AKNS scheme. Condition of commutativity of derivatives defined
in Eqs.~(\ref{eq38}) or their generalizations (see Ref.~\cite{sk-23}) can be used as an
`integrability test': if they commute, then it is quite plausible that the nonlinear wave
equations under consideration are completely integrable and Eqs.~(\ref{eq103}) give useful
information about quasiclassical limit of functions defining the Lax pair in the scalar
representation of the AKNS scheme. At last, the results of the paper can be used in
discussions of problems about propagation of high-frequency wave packets along non-uniform
and evolving with time background, as it was demonstrated in Ref.~\cite{sk-23}.

\begin{acknowledgments}

I thank L.~F.~Calazans de Brito and D.~V.~Shaykin for useful discussions. 
This research is funded by the research project FFUU-2021-0003 of the Institute of Spectroscopy 
of the Russian Academy of Sciences and by the Foundation for
the Advancement of Theoretical Physics and Mathematics ``BASIS''.

\end{acknowledgments}

%\section*{Data Availability Statement}

%The data that support the findings of this study are available
%from the author upon reasonable request. 


\begin{thebibliography}{99}

\bibitem{nmpz-80} V. E. Zakharov, S. V. Manakov, S. P. Novikov, and L. P.
Pitaevskii, {\it The Theory of Solitons: The Inverse Scattering
Method,} (Nauka, Moscow, 1980) (translation: Consultants Bureau, 1984).

\bibitem{as-81} M. J. Ablowitz, H. Segure, {\it Solitons and the Inverse Scattering
Transform,} (SIAM, Philadelphia, 1981).

\bibitem{newell-85} A. C. Newell, {\it Solitons in Mathematics and Physics,} (SIAM, Philadelphia, 1985).

\bibitem{ggkm-67} S.~C.~Gardner, J.~M.~Greene, M.~D.~Kruskal, R.~M.~Miura,
{ Phys. Rev. Lett.,} {\bf 19,} 1095 (1967).

\bibitem{karpman-67} V.~I.~Karpman, { Phys. Lett. A,} {\bf 25,} 708 (1967).

\bibitem{karpman-73} V. I. Karpman, {\it Non-Linear Waves in Dispersive Media,} (Nauka, Moscow, 1973)
(English translation: Pergamon Press, Oxford, 1975).

\bibitem{JLML-99} S. Jin, C. D. Levermore, D. W. McLaughlin,
Comm. Pure Appl. Math., {\bf 52,} 613 (1999).

\bibitem{kku-02} A. M. Kamchatnov, R. A. Kraenkel, B. A. Umarov,
Phys. Rev. E {\bf 66,} 036609 (2002).

\bibitem{zs-73} V. E. Zakharov, A. B. Shabat, Zh. Eksp. Teor. Fiz., {\bf 64,} 1627 (1973)
[Sov. Phys. JETP, {\bf 37,} 823 (1973)].

\bibitem{gp-73} A.~V.~Gurevich and L.~P.~Pitaevskii, Zh. Eksp. Teor. Fiz.,
{\bf 65,} 590 (1973) [Sov. Phys.-JETP, \textbf{38}, 291 (1974)].

\bibitem{whitham} G.~B.~Whitham, Proc. Roy. Soc. London, A \textbf{ 283}, 238 (1965).

\bibitem{Whitham-74} G.~B.~Whitham, {\it Linear and Nonlinear Waves,}
(Wiley Interscience, New York, 1974).

\bibitem{gp-87} A.~V.~Gurevich and L.~P.~Pitaevskii, Zh. Eksp. Teor. Fiz.,
{\bf 93,} 871 (1987) [Sov. Phys.-JETP, \textbf{93}, 871 (1987)].

\bibitem{kamch-21a} A.~M.~Kamchatnov, Usp. Fiz. Nauk., {\bf 191,} 52-87 (2021)
[Phys.--Uspekhi, {\bf 64,} 48-82 (2021)].

\bibitem{lanczos} C.~Lanczos, {\it The Variational Principles of Mechanics,}
(University of Toronto Press, Toronto, 1962).

\bibitem{LL-6} L.~D.~Landau and E.~M.~Lifshitz, {\it Fluid Mechanics,}
(Pergamon, Oxford, 1987).

\bibitem{el-05} G. A. El, Chaos, {\bf 15,} 037103 (2005).

\bibitem{kamch-20a} A.~M.~Kamchatnov, %Theory of quasi-simple dispersive shock waves and number of
%solitons evolved from a nonlinear pulse,
Chaos {\bf 30,} 123148 (2020).

\bibitem{ks-21} A. M. Kamchatnov, D. V. Shaykin, %Propagation of wave packets along intensive simple waves,
Phys. Fluids, {\bf 33,} 052120 (2021).

\bibitem{kamch-19} A.~M.~Kamchatnov, Phys. Rev. E {\bf 99,} 012203 (2019).

\bibitem{kamch-21} A.~M.~Kamchatnov, Zh. Eksp. Teor. Fiz. {\bf 159,} 76 (2021)
[JETP, {\bf 132,} 63 (2021)].

\bibitem{cbk-21} L.~F.~Calazans de Brito, A.~M.~Kamchatnov, %Number of solitons produced from a large initial
%pulse in the generalized NLS dispersive hydrodynamics theory,
Phys. Rev. E {\bf 104,} 054203 (2021).

\bibitem{egkkk-07} G. A. El, A. Gammal, E. G. Khamis, R. A. Kraenkel, A. M. Kamchatnov,
%Theory of optical dispersive shock waves in photorefractive media,
Phys. Rev. A {\bf 76,} 053813 (2007).

\bibitem{egs-08} G.~A.~El, R.~H.~J.~Grimshaw,  N.~F.~Smyth, { Physica D}, {\bf 237,} 2423 (2008).

\bibitem{mfweh-20} M. D. Maiden, N. A. Franco, E. G. Webb, G. A. El, and M. A. Hoefer,
J. Fluid Mech. {\bf 883,} A10 (2020).

\bibitem{poincare} H. Poincar\'{e}, {\it Les m\'ethodes nouvelles de la M\'ecanique c\'eleste,}
t.~III, (Paris, Gauthier-Villar, 1899).

\bibitem{cartan} \'E. Cartan, {\it Le\c{c}ons sur les invariants int\'egraux,} (Hermann, Paris,  1922).

\bibitem{lamb} H.~Lamb, {\it Hydrodynamics,} (Cambridge University Press, Cambridge, 1994).

\bibitem{stokes} G.~G.~Stokes, {\it Mathematical and Physical Papers,} Vol.~V, p.~163
(Cambridge University Press, Cambridge, 1905).

\bibitem{ai-77} O.~Akimoto and K.~Ikeda, J. Phys. A: Math. Gen. {\bf 10,} 425 (1977).

\bibitem{dkn-03} S.~A.~Darmanyan, A.~M.~Kamchatnov, M.~Nevi\'{e}re,
Zh. Eksp. Teor. Fiz., {\bf 123,} 997 (2003) [JETP, {\bf 96,} 876 (2003)].

\bibitem{gant-66} F. R. Gantmacher, {\it Lectures in Analytical Mechanics,} (Moscow, Nauka, 1966)
(English translation: Mir, Moscow, 1975).

\bibitem{LL-3} L.~D.~Landau and E.~M.~Lifshitz, {\it Quantum Mechanics,}
Pergamon, Oxford (1981).

\bibitem{arnold-89} V. I. Arnold, {\it Mathematical Methods of Classical Mechanics,} (N. Y., Springer,1989).

\bibitem{sk-23} D. V. Shaykin, A. M. Kamchatnov, Phys. Fluids (accepted, 2023),
preprint arXiv:2303.16592 (2023).

\bibitem{hoefer-14}  M. A. Hoefer,  { J. Nonlinear Sci.,} {\bf 24,} 525 (2014).

\bibitem{ks-09} A. M. Kamchatnov and M. Salerno, J. Phys. B: At. Mol. Opt. Phys. {\bf 42,} 185303 (2009).

\bibitem{akns-74} M. J. Ablowitz, D. J. Kaup, A. C. Newell, H. Segur, Stud. Appl. Math.
{\bf 53,} 249 (1974).

\bibitem{ak-02} A. M. Kamchatnov and R. A. Kraenkel, J. Phys. A: Math. Gen., {\bf 35,} L13 (2002).

\bibitem{kamch-94} A. M. Kamchatnov, Phys. Lett. A, {\bf 186,}  387 (1994).

\bibitem{kamch-04} A. M. Kamchatnov, Physica D, {\bf 188,}  247 (2004).

\bibitem{ak-01} A. M. Kamchatnov, J. Phys. A: Math. Gen., {\bf 34,} L441 (2001).

\bibitem{krichever-88} I. M. Krichever, Funk. Analiz. Prilozh. {\bf 22,} 37 (1988)
[Funct. Anal. Appl., {\bf 22,} 200 (1988)].

\bibitem{dn-93} B. A. Dubrovin, S. P. Novikov, Sov. Sci. Rev. C. Math. Phys. {\bf 9,} 1 (1993).

\bibitem{alber-93} S. J. Alber, Complex deformations of integrable Hamiltonians over generalized Jacobi varieties,
in {\it Nonlinear Processes in Physics,} eds A. S. Fokas, D. J. Kaup, A. C. Newell, and V. E. Zakharov, p.6 
(Berlin, Springer, 1993). 

%\bibitem{ceh-19} T.~Congy, G.~A.~El, M.~A.~Hoefer, J. Fluid Mech., {\bf 875,} 1145 (2019).









\end{thebibliography}
\end{document}